\pdfoutput=1
\documentclass[runningheads]{llncs}
\usepackage{graphicx}
\usepackage{multirow}
\usepackage{tabularx}
\usepackage{listings}
\usepackage{color}
\usepackage{cite}
\usepackage{algorithm}
\usepackage{booktabs}
\usepackage[noend]{algpseudocode}

\newcolumntype{Y}{>{\centering\arraybackslash}X}
\newcolumntype{Z}{p{4cm}>{\centering\arraybackslash}X}
\newcolumntype{b}{>{\small}X}
\newcolumntype{s}{>{\hsize=.2\hsize}Y}

\definecolor{backcolour}{rgb}{0.9,0.9,0.9}
\definecolor{purple}{rgb}{0.65, 0.12, 0.82}
\definecolor{darkgray}{rgb}{.4,.4,.4}

\lstdefinestyle{mystyle}{
    keywords={returns, bool, uint256, true, false, ether},
    keywordstyle=\color{blue}\bfseries,
    ndkeywords={private, view, block, abi},
    ndkeywordstyle=\color{darkgray}\bfseries,
    comment=[l]{//},
    commentstyle=\color{purple}\ttfamily,
    backgroundcolor=\color{backcolour},
    basicstyle=\small,
    breakatwhitespace=false,         
    breaklines=true,                 
    captionpos=b,                    
    keepspaces=true,                 
    numbers=left,                    
    numbersep=5pt,                  
    showspaces=false,                
    showstringspaces=false,
    showtabs=false,                  
    tabsize=2
}
 
\begin{document}

\title{Characterizing Code Clones in the Ethereum Smart Contract Ecosystem}
\vspace{-0.1in}

\author{Ningyu He\inst{1} \and
Lei Wu\inst{2}\and
Haoyu Wang\inst{1}\and
Yao Guo\inst{3}\and
Xuxian Jiang\inst{2}}

\authorrunning{N. He et al.}

\institute{Beijing University of Posts and Telecommunications \and
PeckShield, Inc. \and
Peking University}

\maketitle

\vspace{-0.2in}
\begin{abstract}

In this paper, we present the first large-scale and systematic study to characterize the code reuse practice in the Ethereum smart contract ecosystem. We first performed a detailed similarity comparison study on a dataset of 10 million contracts we had harvested, and then we further conducted a qualitative analysis to characterize the diversity of the ecosystem, understand the correlation between code reuse and vulnerabilities, and detect the plagiarist DApps. Our analysis revealed that over 96\% of the contracts had duplicates, while a large number of them were similar, which suggests that the ecosystem is highly homogeneous. Our results also suggested that roughly 9.7\% of the similar contract pairs have exactly the same vulnerabilities, which we assume were introduced by code clones. In addition, we identified 41 DApps clusters, involving 73 plagiarized DApps which had caused huge financial loss to the original creators, accounting for 1/3 of the original market volume.

\vspace{-0.1in}
\end{abstract}

\section{INTRODUCTION}
\vspace{-0.1in}
With the widespread application of blockchain techniques, cryptocurrency has experienced an explosive growth during the last decade. As one of the most revolutionary and representative platforms after Bitcoin~\cite{bitcoin}, Ethereum~\cite{ethereum} has attracted a large number of participants, including developers and users, and becomes one of the most active communities in the cryptocurrency world.

Smart contract, the most important innovation of Ethereum, provides the ability to ``digitally facilitate, verify, and enforce the negotiation or performance of a contract''~\cite{wiki:ethereum}, while the correctness of its execution is ensured by the consensus protocol of Ethereum. Such a courageous attempt has been approved by the market, \textit{i.e}., Ethereum's market cap was around $\$14.5$B on February 26th, 2019~\cite{coinmarketcap}, the largest volume besides Bitcoin. As of this writing, roughly 10 million smart contracts have been deployed on the Ethereum Mainnet. 

Smart contracts are typically written in higher level languages, \textit{e.g.}, Solidity~\cite{solidity} (a language similar to JavaScript and C++), then compiled to EVM bytecode. As one of the most important rules on Ethereum, ``Code is Law'' means all executions and transactions are final and immutable.

As a result, one main characteristic of smart contracts is that a considerable number of them published source code to gain the users' trust and prove the security of their code, especially for the popular ones~\cite{etherscan}. This feature is more noticeable for Decentralized Applications (DApps for short, which consist of one or more contracts). In general, its code base should be available for scrutiny and it should be governed by autonomy, different from the traditional closed-source applications that require the end users to trust the developers in terms of decentralization as they cannot directly access data via any central source.

However, the open-source nature of smart contracts has provided convenience for plagiarists to create \textbf{contract clones}, \textit{i.e.}, copying code from other available contracts. 
The impacts of contract clones are mainly of twofolds.
On one hand, the plagiarists could insert arbitrary/malicious code into the normal contracts. A typical example is the so-called \texttt{honeypot smart contracts}~\cite{honeypot1, honeypot2, honeypot3}, which are scam contracts that try to fool users with stealthy tricks. 
On the other hand, as many smart contracts are suffering from serious vulnerabilities, the copy-paste vulnerabilities would be inherited by the plagiarized contracts.

\begin{figure}[t]
\centering
\includegraphics[width=0.9\textwidth]{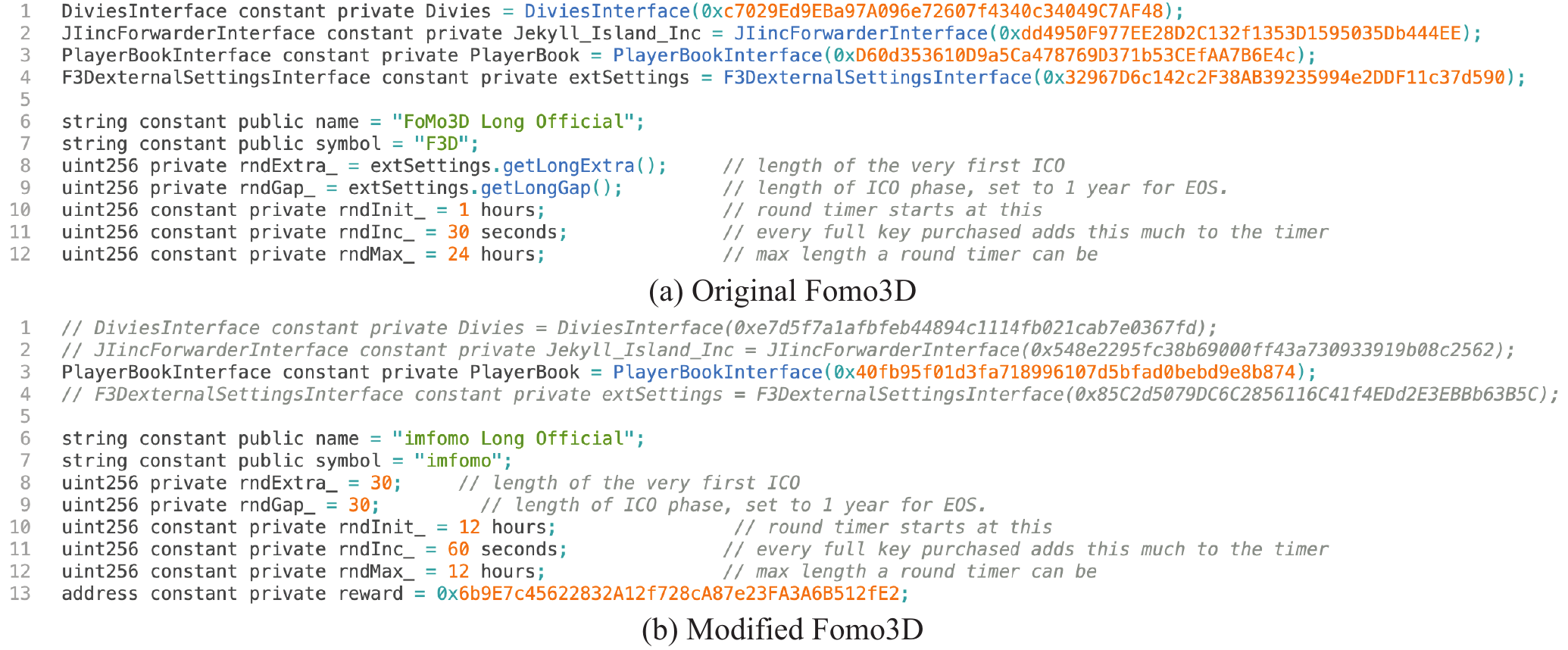}
\vspace{-0.15in}
\caption{The original Fomo3D and a plagiarized contract from it.
}
\vspace{-0.3in}

\label{fig:fomo3dplagiarism}
\end{figure}

Here, we use Fomo3D~\cite{fomo3d} as a motivating example, which is a popular and phenomenal Ponzi-like game. At its peak in 2018, Fomo3D had over $10,000$ daily active users with a volume of over $40,000$ ETHs~\cite{officialfomo3d}. As a result, numerous Fomo3D-like games sprang up with plagiarism behaviors by simply reusing the source code of the original one. Unfortunately, some hackers had figured out the design flaw of the airdrop mechanism in the original Fomo3D~\cite{fomo3dattack}. 
Consequently, almost all the awkward imitators were exposed to those attackers. \textit{LastWinner}, one of the most successful followers of Fomo3D, was attacked and lost more than $5,000$ ETHs within $4$ days~\cite{lastwinnerattacked}. 
Figure~\ref{fig:fomo3dplagiarism} shows a plagiarized contract example originated from Fomo3D~\cite{officialfomo3d}. Interestingly, the vulnerable part was kept wholly intact by the plagiarist, but all the dependent contracts, and some arguments like round timer and round increment, were modified to make it appear as a brand new game as shown in Figure~\ref{fig:fomo3dplagiarism}. 

In this paper, we present a large-scale systematically study to characterize the code clone behaviors of Ethereum smart contracts in a comprehensive manner. 
To this end, we have collected by far the largest Ethereum smart contract dataset with nearly 10 million smart contracts, being deployed from July 2015 to December 2018. 
To address the scalability issues introduced by the large scale dataset, we first seek to identify the duplicate contracts by removing function unrelated code (\textit{e.g.}, creation code and Swarm code), and tokenizing the code to keep opcodes only. 
After the pre-processing step to remove duplicate contracts, the dataset has been shrunk to less than 1\% of the original size. 
For the remaining $78,611$ distinct contracts, we take advantage of a customized fuzzy hashing approach to generate the fingerprints and then conduct a pair-wise similarity comparison. 
Specifically, we adopt a pruning strategy to discard ``very different'' contracts by comparing the meta features (\textit{e.g.}, length of opcode), to accelerate the comparison procedure. 
Based on a similarity threshold of 70, we are able to identify $472,663$ similar smart contract pairs (with $47,242$ contracts involved) for user-created contracts, which suggested that over $63.29\%$ of the distinct user-created contracts have at least one similar contract in our dataset.

The preliminary exploration has identified a huge number of duplicates and many similar contracts as well. 
Then, we further seek to understand the reasons leading to contract clones and characterize their security impacts.

\noindent \textbf{(1) The reasons leading to contract clones.} 
    We further group the smart contracts into clusters.
    Over 60\% of the distinct contracts were grouped into roughly 10K clusters, while the cluster distribution follows a typical Pareto effect.
    Top 20\% of the clusters occupied over 60\% of the distinct contracts. With regard to the whole dataset including all the duplicates, the top 1\% of the clusters account for 95\% of the contracts.
    ERC20 token contracts, ICO and AirDrop, and Game contracts are the most popular clusters. 
    A large number of similar contracts were created based on the same template (\textit{e.g.}, ERC20 template). 
    We have manually summarized a list of 53 different common templates used in the Ethereum smart contract ecosystem.
    \textbf{This result reveals the homogeneity nature of the smart contract ecosystem.}
    
    \noindent \textbf{(2) Vulnerability Provenance.} Copy-paste vulnerabilities were prevalent in most popular software systems. 
    Here, we study the relationship between contract clones and the presence of vulnerabilities, from two aspects. 
    First, we scanned all the unique contracts using a state-of-the-art vulnerability scanner~\cite{peckshieldscanner}. Over $20,346$ district smart contracts ($27.26\%$) contain at least one vulnerability. \textbf{Considering the large number of duplicates, we were able to identify ten folds of vulnerable contracts ($205,010$)}.
    Then, for the distinct contracts, as a number of them have similar code, we further compare whether they were exposed to similar vulnerabilities. \textbf{Overall, our results suggest that roughly 9.7\% of the similar contract pairs have exactly the same vulnerabilities, which we assume were introduced by code clones.}
    
    \noindent \textbf{(3) Plagiarized DApps.} 
    As a DApp is more complicated than smart contract, \textit{i.e.}, one DApp could include one or more contracts, thus we further study the similarity between DApps, seeking to identify the plagiarized ones. Using a bipartite graph matching approach, we identified 41 DApp clusters, involving 73 plagiarized DApps. \textbf{The plagiarized ones have caused huge financial loss to the original creators, accounting for 1/3 of the original volume.}

To the best of our knowledge, this is the first systematic study of code clones in the Ethereum smart contract ecosystem \emph{at scale}. Our results revealed the highly homogeneous nature of the ecosystem, \textit{i.e.}, code clones are prevalent, while it greatly helps spread vulnerabilities and ease the job of plagiarists.
Our results motivated the need for research efforts to identify security issues introduced by copy-paste behaviors. 
Our efforts can positively contribute to the smart contract ecosystem, and promote the best operational practices for developers.

\vspace{-0.1in}
\section{BACKGROUND}
\label{sec:background}

\vspace{-0.1in}
\subsection{Ethereum}
\vspace{-0.1in}

\textbf{External Owned Account vs. Contract Account.}
The basic unit of Ethereum is an \textit{account} and there are two types of accounts~\cite{whitepaper}: External Owned Account (EOA) and Contract Account. 
An EOA is controlled by private keys that are externally owned by a user. More importantly, there is no code associated with it. One can send messages from an EOA by creating and signing a transaction. On the contrary, a contract account is controlled by its associated contract code, which might be activated on receiving a message.

\noindent \textbf{User-created Contract vs. Contract-created Contract.}
An Ethereum smart contract can be created either by a user, or by another existing contract. In this paper, we follow the terminology ``user-created contract'' and ``contract-created contract'' adopted by~\cite{kiffer2018analyzing} to distinguish these two types of creations.

\noindent \textbf{Decentralized applications (DApps).}
Ethereum aims to create an alternative protocol to build DApps~\cite{dappdefine}, which are stored on and executed by the Ethereum system. Specifically, a DApp is \textit{a contract or a collection of contracts that have an interface on the Internet, typically a website or a browser game, which could be interacted by players or users directly}. A number of websites were emerged to host the list of DApps~\cite{dapptotal, dappreview, dappradar}.

\begin{figure}[t]
\includegraphics[width=0.9\textwidth]{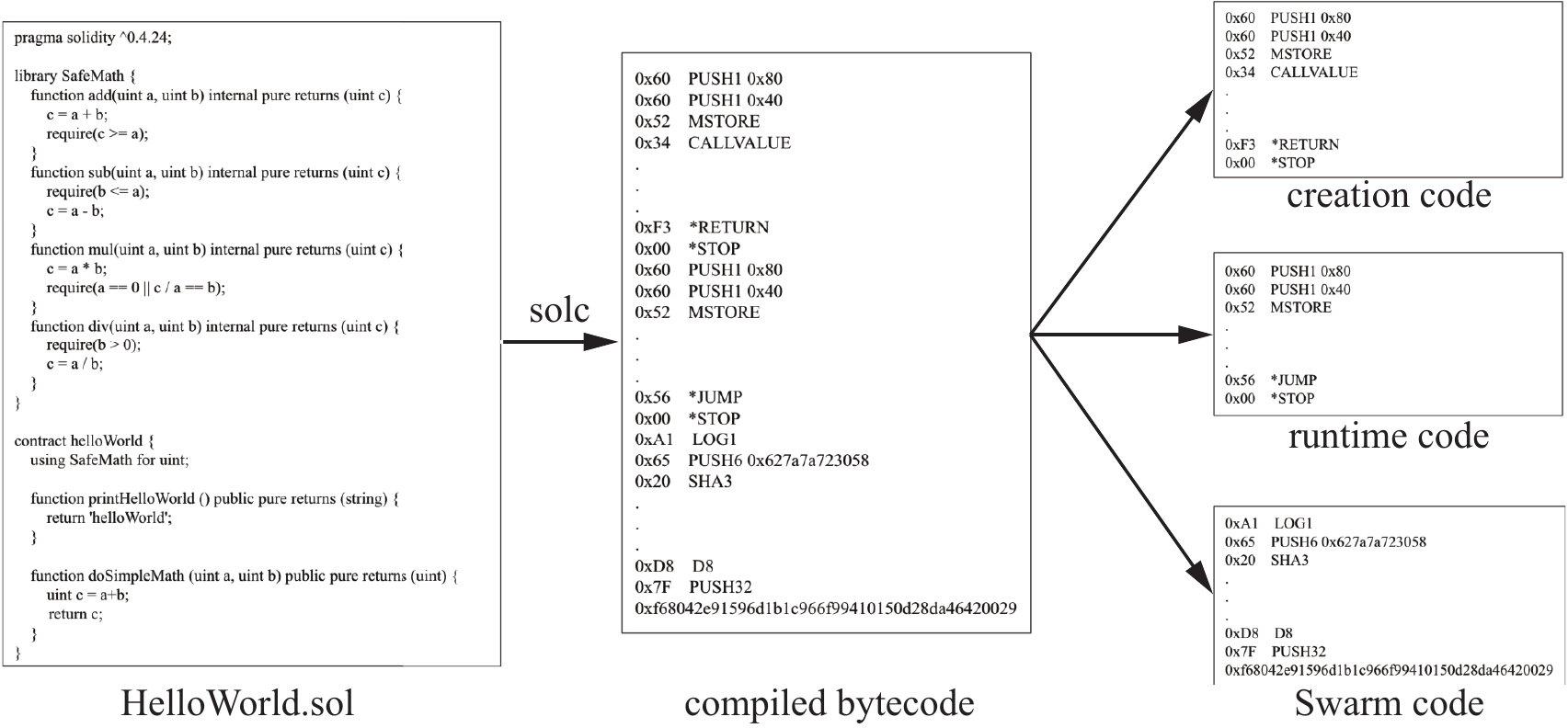}
\vspace{-0.15in}
\caption{An example of Helloworld.sol and its corresponding bytecode.
} 
\vspace{-0.2in}
\label{fig:fig2}
\end{figure}

\vspace{-0.1in}
\subsection{Ethereum Virtual Machine (EVM)}
\vspace{-0.1in}
EVM is the runtime environment for smart contracts in Ethereum. Specifically, a sandboxed virtual stack machine is embedded within each full Ethereum node, responsible for executing contract bytecode with a 256-bit register stack~\cite{evm}.
EVM is not as complex as traditional operating systems. Its operators and operands are all pushed onto the stack indistinguishably, except for data that require persistent storage space on Ethereum. Therefore, all the immediate numbers and data to be used by the operation code will be \textit{pushed} onto the stack.

Generally, developers implement their smart contracts with the Solidity language, then build the source code using the Solidity compiler, a.k.a. \textit{solc}, to generate the EVM bytecode. A typical EVM bytecode is composed of three parts: \textit{creation code}, \textit{runtime code} and \textit{swarm code}, as shown in Figure~\ref{fig:fig2}.

\textbf{Creation code} is only executed by EVM once during the transaction of the contract deployment. It determines the initial states of the smart contract being deployed and returns a copy of the runtime code. A typical creation code ends with the following operation sequence: \texttt{PUSH 0x00, RETURN, STOP}, \texttt{0x6000f300} in bytecode, as shown in Figure~\ref{fig:fig2}.

\textbf{Runtime code} is the most crucial part, including function selector, function wrapper, function body and exception handling. Based on the corresponding operations, EVM will execute runtime code accordingly. Besides, in order to label jumping destinations of the function selector, solc sorts functions by their signatures, \textit{i.e.}, the leading 4 bytes of the SHA-3 hashes of function declarations with a well-defined format~\cite{funcsig}.
Accordingly, adding new functions or deleting existing ones will not affect the relative order of the remaining functions.

\textbf{Swarm code} is not served for execution purpose. Solc uses the metadata of a contract, including compiler version, source code and the located block number, to calculate the so-called Swarm hash, which can be used to query on \textit{Swarm}, a decentralized storage system, to prove the consistency between the contract you see and the contract being deployed, namely \textit{what you see is what you get}. As a result, re-deploying a smart contract would result in a different swarm code, even with the same creation code and runtime code. Swarm code always begins with \texttt{0xa165}, \textit{i.e.}, \texttt{LOG1 PUSH 6}. The following six bytes are \texttt{0x627a7a723058}, whose leading four bytes can be decoded as ``bzzr'', the Swarm's URL scheme. Furthermore, Swarm code always ends with \texttt{0x0029}, which means the hash part length between \texttt{0xa165} and \texttt{0x0029} is 41 bytes long. In short, we are able to identify the swarm code quickly and precisely based on those hard-coded bytes.

\vspace{-0.1in}
\section{Methodology}
\label{sec:methodology}
\begin{figure}[t]
\centering
\includegraphics[width=0.85\textwidth]{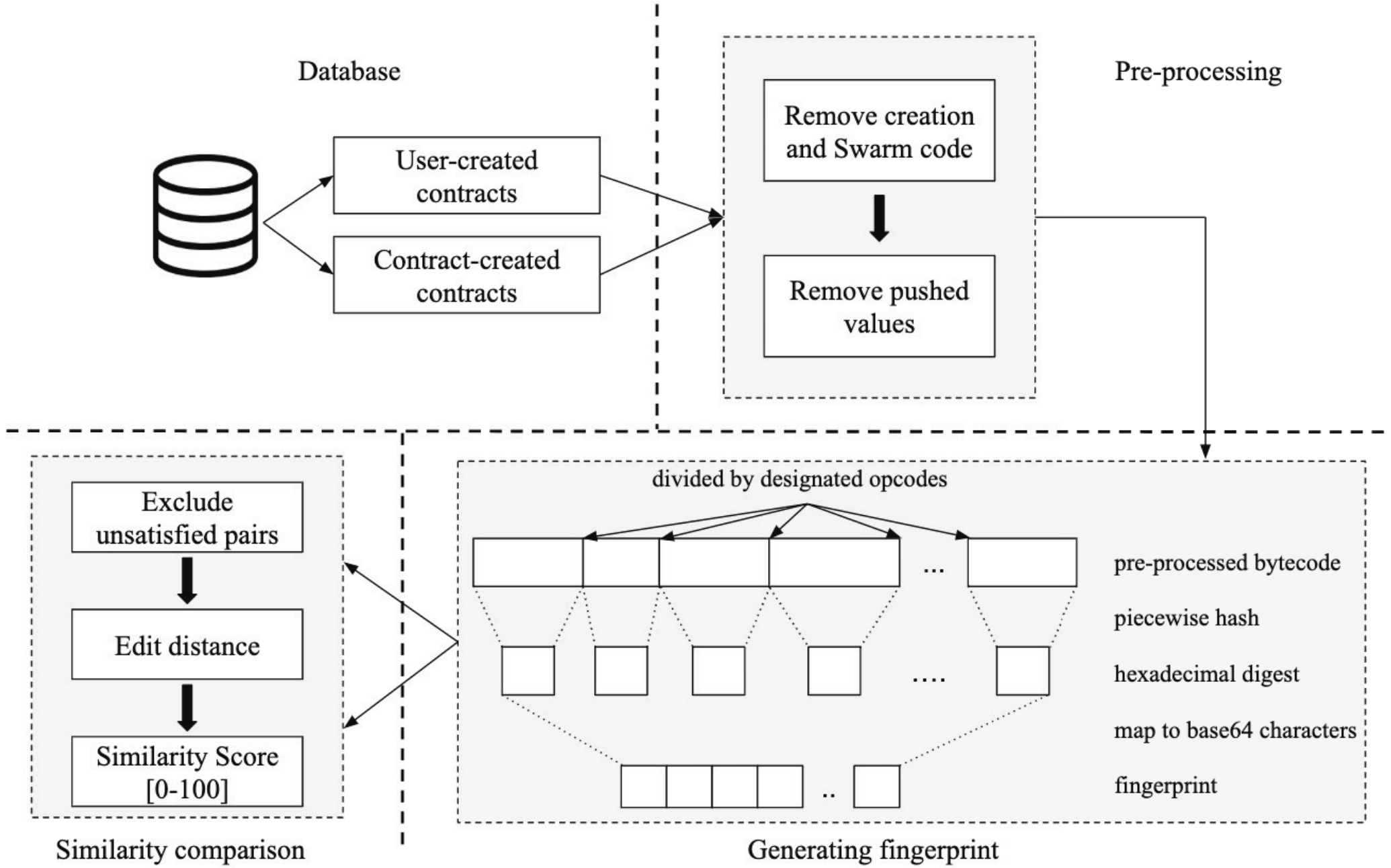}
\vspace{-0.15in}
\caption{An overview of our approach on smart contract similarity comparison.} 
\vspace{-0.2in}
\label{fig:methodology}
\end{figure}

\vspace{-0.1in}
\textbf{Overall Process.}
We summarize our approach in Figure~\ref{fig:methodology}. The pipeline starts with the dataset with nearly 10 million smart contracts we have collected. 
We first seek to remove duplicate smart contracts to reduce the computational workload in two steps: 1) removing the useless creation and Swarm code parts; 2) removing all assigned values in assignment statements and function calls for desensitization. To this end, the smart contracts were scanned for tokenization by generating token hashes, which allow us to capture subtle differences of the clones.
After that, for the remaining contracts with distinct token hashes, we take advantage of a customized fuzzy hashing approach to generate the fingerprints. Lastly, we enforce a pair-wise comparison strategy with pruning to achieve scalability. The output of the whole analysis pipeline is a set of contract clone pairs with the corresponding similarity scores. 

Note that the output result will be further correlated with our in-depth analysis in Section~\ref{sec:qualitative}, including contract clustering, vulnerability provenance and DApps plagiarism detection.

\vspace{-0.1in}
\subsection{Pre-processing}
\label{sec:pre-processing}
\vspace{-0.1in}

The purpose of pre-processing is of two folds: first identifying the duplicate contracts, and then tokenizing contracts for further comparison.

As we mentioned, creation code and Swarm code have nothing to do with  similarity calculation. Fortunately, they can be easily identified and removed from the bytecode. Afterwards, we use a hash set to guarantee the uniqueness of the remaining contracts in terms of runtime code.
Secondly, to enable fast and accurate fingerprint generation, we further remove all the immediate numbers after opcode \texttt{PUSH} since EVM is a stack-based virtual machine. Again, we use a hash table to guarantee the uniqueness of the remaining contracts. In this way, we obtain a minimized database with little feature lost for similarity detection.

\vspace{-0.1in}
\subsection{Generating Fingerprint}
\label{sec:generating-fingerprint}
\vspace{-0.1in}

Calculating the edit distance between two given sequences is a well-known way to measure their similarity. 
In this work, we use a \textit{fuzzy hashing} technique~\cite{fuzzyhashing} to condense the original bytecode to a much shorter fingerprint and then calculate the edit distance between two fingerprints.
Unlike traditional hash functions, fuzzy hashing first divides the bytecode sequence into smaller pieces, then uses a piece-wise hash function to perform the calculation for each piece and finally concatenates those generated piece-wise hashes to form a fingerprint. 
Suppose someone modifies one particular function, all the related pieces would generate different piece-wise hashes with the original ones, but the other pieces were not affected at all.
In short, fuzzy hashing has advantages of accurate representation and less computing-time consumption.

However, there still exists challenges to determine the boundary of each piece. Previous work chooses a boundary randomly or simply divides the sequence by a pre-defined step (\textit{e.g.}, seven bytes)~\cite{andrew:spamsum}.
Nevertheless, a smart contract is not just a piece of plain-text, and definitely has its semantic meaning. To address the problem, we propose a customized fuzzy hashing algorithm, which is capable of segmenting smart contracts precisely to generate feasible piece-wise hashes for further analysis.

\begin{figure}[t]
\centering
\includegraphics[width=0.8\textwidth]{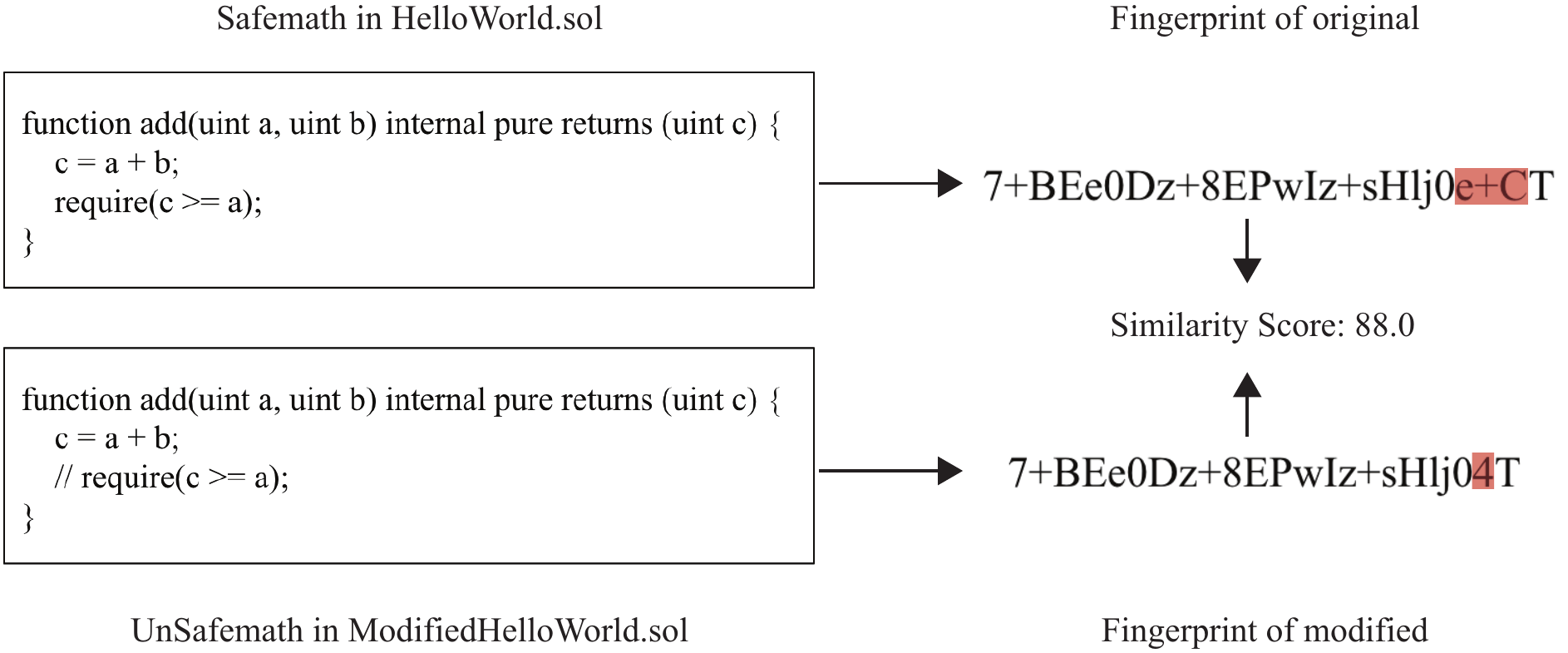}
\vspace{-0.2in}
\caption{An example of fingerprint generation and similarity comparison.
} 
\vspace{-0.2in}
\label{fig:fingerprint}
\end{figure}

\noindent \textbf{Customized Fuzzy Hashing.}
After investigating the bytecode and its execution procedure in EVM, we identify the runtime code that can be further divided into several sub-sequences to perform a basic block level analysis. In Solidity, opcodes \texttt{JUMP}, \texttt{JUMPI}, \texttt{REVERT}, \texttt{STOP}, \texttt{RETURN} are the indicators of the interruption of logical relationship, and these opcodes often mean that the current block should be terminated in building the control flow graph (CFG). Futhermore, as we mentioned in Section~\ref{sec:background}, runtime code always keeps the order of function selector, function wrapper, \textit{etc.}, and maintains the relative order between functions.
After dividing, the piece-wise hash function will be applied on each of the blocks to generate a four byte hexadecimal digest and then mapped to a base-64 character after modulo 64. Finally, a fingerprint is generated by concatenating these characters. The algorithm and details are described in Algorithm~\ref{algo:fingerprintalgo} (cf. Appendices) and a concrete example is given in Figure~\ref{fig:fingerprint}.

\vspace{-0.1in}
\subsection{Similarity Comparison}
\label{sec:similarity-comparision}
\vspace{-0.1in}

At this stage, we are able to perform pair-wise comparison to characterize the similarity between contracts. Since pair-wise comparisons are computationally expensive (billions of comparisons), we propose a pruning strategy here to tackle the problem. 
Intuitively, similar contracts should share similar attributes with minor modifications, opcode length in particular. 
If two contracts are ``very different'' in the opcode length, we will stop comparing the fingerprints and mark them as dissimilar. 
In practice, if more than 30\% attributes of two smart contracts are different, the comparison process will stop.

For each contract pair, we calculate the edit distance between the fingerprints, and then map it to a similarity score in the range of 0 to 100, as follows:

\vspace{-0.1in}
\begin{equation}
\label{formula:simscore}
similarity Score =\left[1-\frac{distance}{\max (len(fp1), len(fp2))}\right] * 100
\end{equation}

Figure~\ref{fig:fingerprint} shows an example of the fingerprints we generated for \texttt{HelloWorld.sol} and its modified version with a slight change, respectively.  
In the modified version, we have removed the \textit{require} statement of the \textit{SafeMath} library, which may lead to an overflow vulnerability. 
The difference between these two fingerprints is highlighted. Obviously, only a few characters within the fingerprint have changed, and the similarity score calculated by our approach is $88.0$.

\vspace{-0.1in}
\section{QUANTITATIVE ANALYSIS}
\vspace{-0.1in}

In this section, we focus exclusively on quantitative analysis, which provides some straightforward but interesting findings we observed before we perform more detailed analysis in Section~\ref{sec:qualitative}. 
\vspace{-0.2in}
\subsection{Dataset}
\vspace{-0.1in}

We have collected by far the largest smart contract dataset, covering almost 10 million smart contracts deployed on the Ehtereum mainnet from July 30th, 2015 to December 31st, 2018. 
As shown in Table~\ref{amount}, only 2.1 million contracts are user-created, and the number of contract-created contracts is four times greater than  user-created ones. 
They were owned by $124,015$ accounts, including $94,307$ for the user-created contracts and $29,708$ for the contract-created contracts.

\begin{table}[t]
\caption{An overview of the dataset before and after pre-processing.}
\vspace{-0.1in}
\label{amount}
\begin{tabularx}{\textwidth}{r|YYY}
\hline
Contract type & \# Contracts (\# Owned Accounts) & After Swarm code removing & After push arguments removing \\
\hline\hline
user-created & 2,121,745 (94,307) & 105,258 & 74,647 \\
contract-created & 7,729,012 (29,708) & 4,539 & 3,964 \\
\hline
\end{tabularx}
\end{table}
\vspace{-0.2in}

\subsection{Pre-processing}
\vspace{-0.1in}

The pre-processing step is helpful in removing duplicates. It turns out that the proportion of contracts to be analyzed has been shrunk dramatically to $0.798\%(78,611/9,850,757)$ of the original dataset we collected. Especially for the contract-created contracts, only $3,964$ distinct contracts remained. 

\begin{figure}[t]
\includegraphics[width=\textwidth]{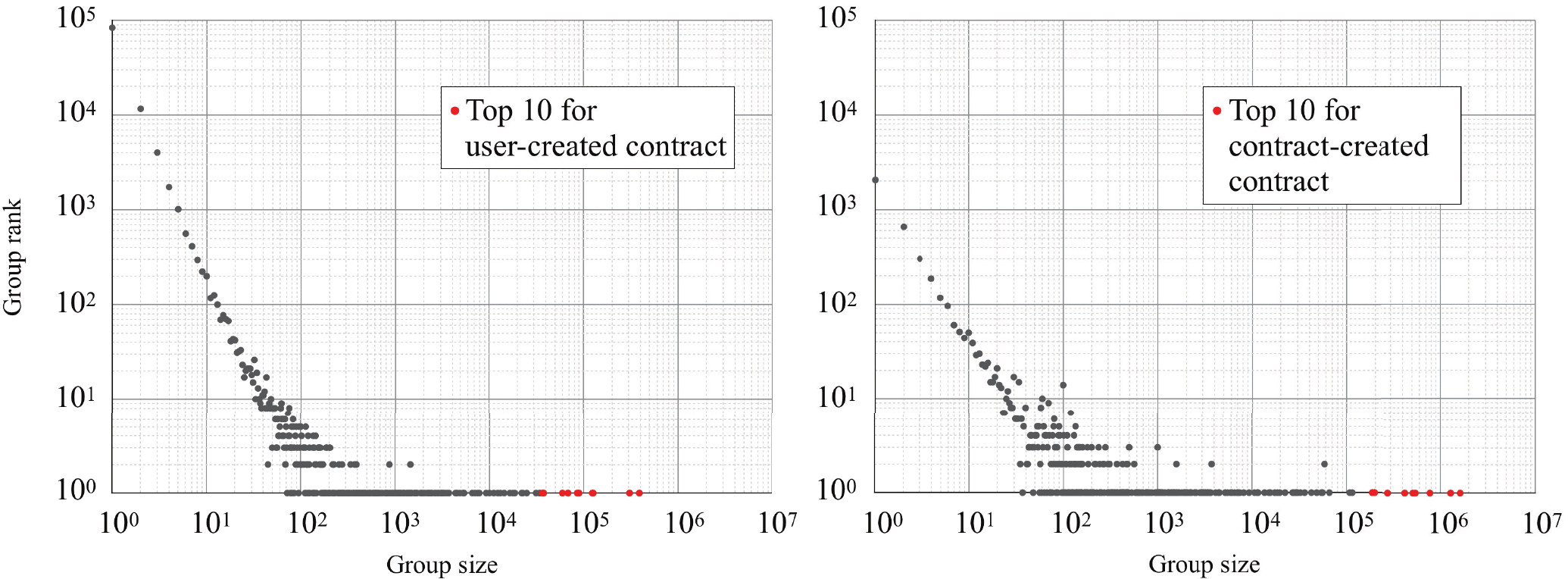}
\vspace{-0.3in}
\caption{The distribution of contract clusters grouped by opcode hash values.} 
\vspace{-0.1in}
\label{fig:groupsizeandrank}
\end{figure}

To figure out the reason for the huge number of duplicates, we first
grouped the duplicated contracts into clusters, and then analyzed the distribution of those clusters, as shown in Figure~\ref{fig:groupsizeandrank}.
We also list the top 10 contracts with the most duplicates in Table~\ref{top10}. It shows that the top 10 clusters occupy the majority of contracts, which represent $62.37\%$ of all user-created contracts and $82.26\%$ of all contract-created contracts, respectively.

After further investigation, we found that most of the user-created contracts belong to a type of wallet, which was named as \texttt{transfer wallet} in the following context.
The transfer wallet can be used in different ways, like avoiding regulation by initiating multiple and multilevel small transfers or reducing risk by splitting a large balance from one account to several small accounts.
Some contracts are regarded as \texttt{forwarder}s (the contract name), which are not wallets but might be functionally similar to those transfer wallets in some way, such as transferring ETHs or tokens. Note that there are forwarders in contract-created contracts as well.
Besides, some of the other duplicated contracts are controlled by exchanges, \textit{e.g.}, Poloniex~\cite{poloniex}, to manage issued tokens, such as Golem~\cite{golem} and Storj~\cite{storj}.

As for the contract-created contracts, some clusters are owned by the Bittrex exchange~\cite{bittrex}. More interestingly, the second largest cluster is a token issued by Gastoken~\cite{gastoken}, which allows users to make profits by tokenizing gas based on the refund mechanism on storage in Ethereum. 
We also found lots of \texttt{Proxy} contracts, which were used to redirect all incoming message calls to other deployed contracts. 
In addition, many contracts belong to ENS~\cite{ens} (Ethereum Name Service), a naming system based on the Ethereum Blockchain. 
Finally, there are two interesting groups related to CryptoMidwives~\cite{cryptomidwives}, which are a kind of contracts aiming to get profit from `CryptoServices' (\textit{i.e.},  CryptoKitties-like games).

\begin{table}[t]
\caption{The top 10 contracts with the most number of duplicates.}
\vspace{-0.1in}
\label{top10}
\begin{tabularx}{\textwidth}{cccc}
\hline
\multicolumn{2}{c}{User-created contracts} & \multicolumn{2}{c}{Contract-created contracts} \\
\# Duplicates & Use & \# Duplicates & Use \\
\hline\hline
390,020 & Transfer wallet & 1,619,511 & Bittrex wallet \\
306,600 & Transfer wallet & 1,284,440 & Gastoken \\
125,929 & Transfer wallet & 776,441 & Bittrex wallet \\
123,787 & Transfer wallet & 544,834 & Proxy \\
89,134 & Transfer wallet & 540,094 & Proxy \\
85,782 & Transfer wallet & 511,894 & Forwarder \\
68,297 & Token manager of Poloniex & 420,822 & ENS \\
59,543 & Token-only forwarder & 277,380 & CryptoMidwives \\
37,628 & Transfer wallet & 196,260 & CryptoMidwives \\
36,625 & Token manager of Poloniex & 185,889 & Forwarder \\
\hline
\end{tabularx}
\vspace{-0.1in}
\end{table}

\vspace{-0.1in}
\subsection{Similarity Comparison}
\vspace{-0.1in}

For all the original 10 million smart contracts, it would be unfeasible for us to perform pair-wise comparison.
Taking advantage of our pruning strategies, we are able to narrow down the contract pairs by almost four orders of magnitude, which greatly reduces the burden on similarity comparison. 

\begin{figure}[t]
\includegraphics[width=\textwidth]{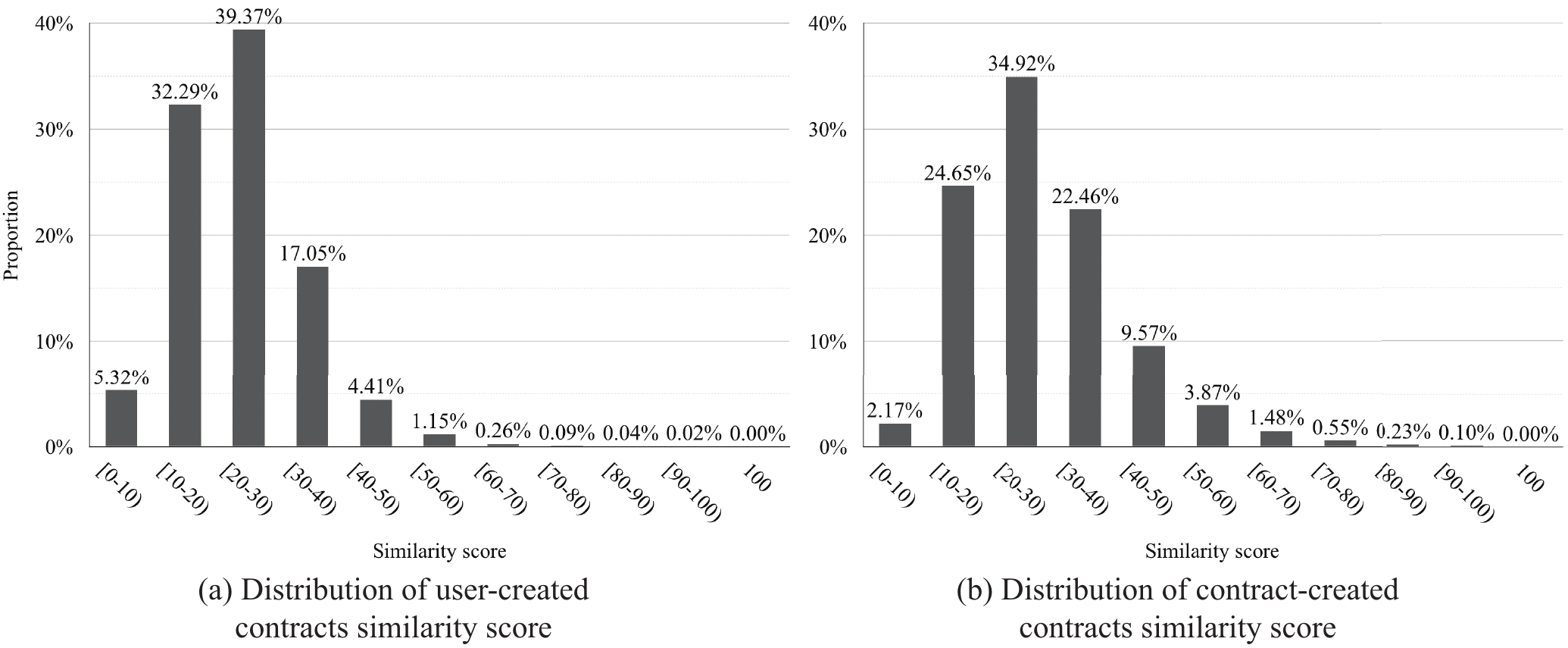}
\vspace{-0.3in}\caption{The distribution of similarity scores for smart contract pairs.
} \vspace{-0.2in}
\label{fig:distribution_similarity}
\end{figure}

\noindent \textbf{The distribution of similarity score.} 
With our pruning strategies, over 308 million user-created contract pairs and 1.2 million contract-created pairs were compared, and Figure~\ref{fig:distribution_similarity} shows the distribution of similarity scores. 
Roughly 90\% of the contract pairs have similarity scores less than 40, while only a small percentage of contract pairs have similarity scores higher than 70, among them are 0.153\% of user-created contracts and 0.879\% of contract-created contracts. In addition, 300 contract pairs have exactly the same similarity scores.

We further performed a comprehensive investigation of the contract pairs at different similarity ranges. We randomly sampled 100 pairs at different ranges and manually examined their source code and bytecode. We concluded that $70$ is the best threshold, \textit{i.e.}, a contract pair with a similarity score higher than $70$ will be regarded as the similar pair, which is in line with other fuzzy hashing based code clone detection studies~\cite{zhou2012detecting}. 
In the end, $472,663$ user-created contracts pairs (with 47,242 contracts involved) and $11,161$ contract-created contracts pairs (with 2,409 contracts involved) were considered as similar pairs. 

\vspace{-0.1in}
\section{Qualitative Analysis}
\label{sec:qualitative}
\vspace{-0.1in}

Our previous observations suggest 
that over $96.07\%$ of user-created contracts and $99.97\%$ of contract-created contracts have duplicates, and a large number of contract pairs were similar. 
In this section, we delve deeper into qualitative evaluation. 
We first seek to cluster the distinct contracts into groups based on their similarity scores, for which we try to understand the reasons leading to contract clones and study the diversity of the ecosystem (\textit{e.g.}, what are these contracts?).
Then we propose to explore the correlation between code clone and vulnerabilities, \textit{i.e.}, whether code clones lead to the spread of vulnerabilities. 
At last, we try to identify the DApp Clones in the wild and measure their impact.

\vspace{-0.1in}
\subsection{Clustering Smart Contracts}
\label{subsec:cluster}
\vspace{-0.1in}

\noindent \textbf{The Clustering Approach.}
Here, we use a simple but effective approach to cluster these contracts based on their similarity scores. 
Specifically, we cluster any contract pair whose similarity score is $70$ or higher. 
Therefore, we are able to build a \texttt{weighted undirected graph} by treating each contract as a node. There will be an edge between two nodes if their similarity score (i.e., weight) is larger than or equal to $70$.
Then, we traverse the graph and consider each \texttt{connected component} as a cluster.
To sum up, only unique contracts are used to construct the graph, and only contracts with edges whose weights are higher than $70$ can be regarded as a connected component to form a cluster.

\begin{figure}[t]
\centering
\includegraphics[width=0.9\textwidth]{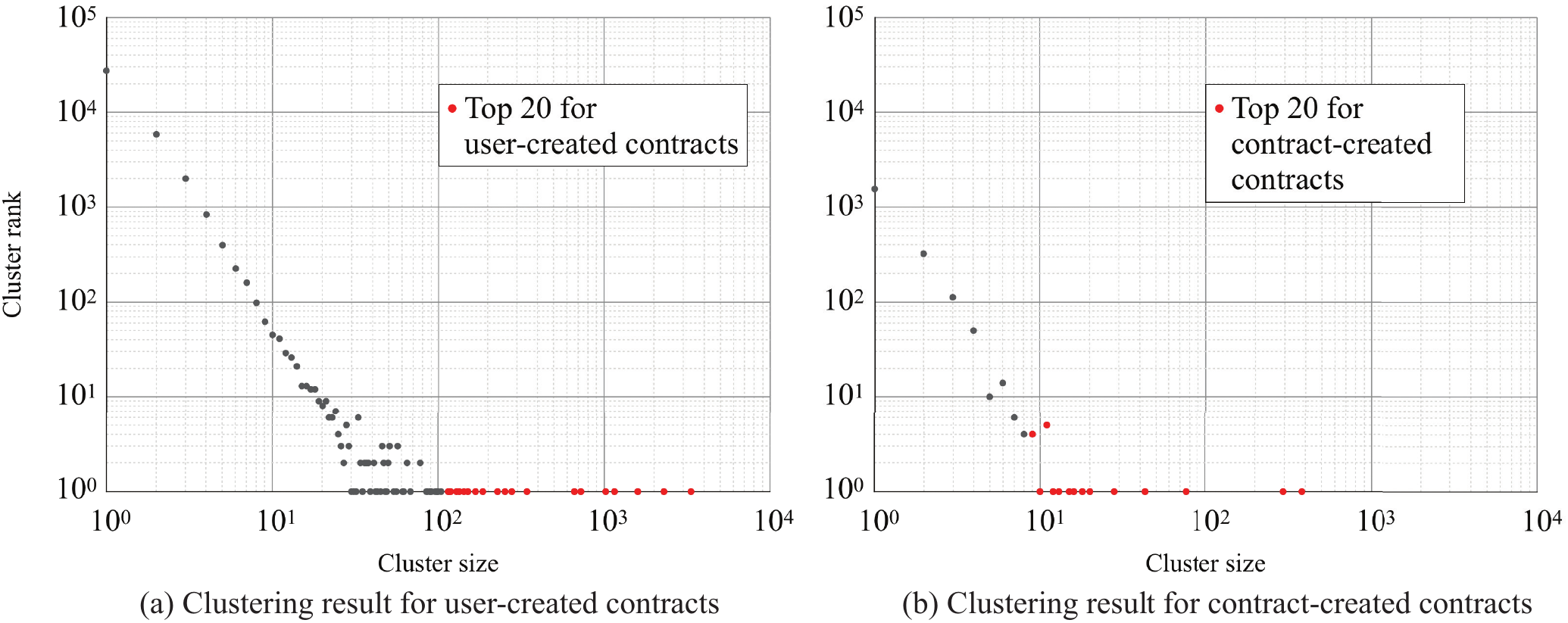}
\vspace{-0.15in}
\caption{The distribution of clusters.
} 
\vspace{-0.1in}
\label{fig:cluster}
\end{figure}

\noindent \textbf{Clustering Result.}
We apply the clustering approach on user-created contracts and contract-created contracts, respectively. The results are presented in Figure~\ref{fig:cluster}.
Each of them follows a long-tail distribution.
For user-created contracts, over $63.29\%$ of them were clustered into 9,971 clusters, while 27,405 of them were isolated nodes in the graph.
For contract-created contracts, $60.77\%$ of them were clustered into 2,409 clusters.

\begin{figure}[t]
\centering
\includegraphics[width=0.9\textwidth]{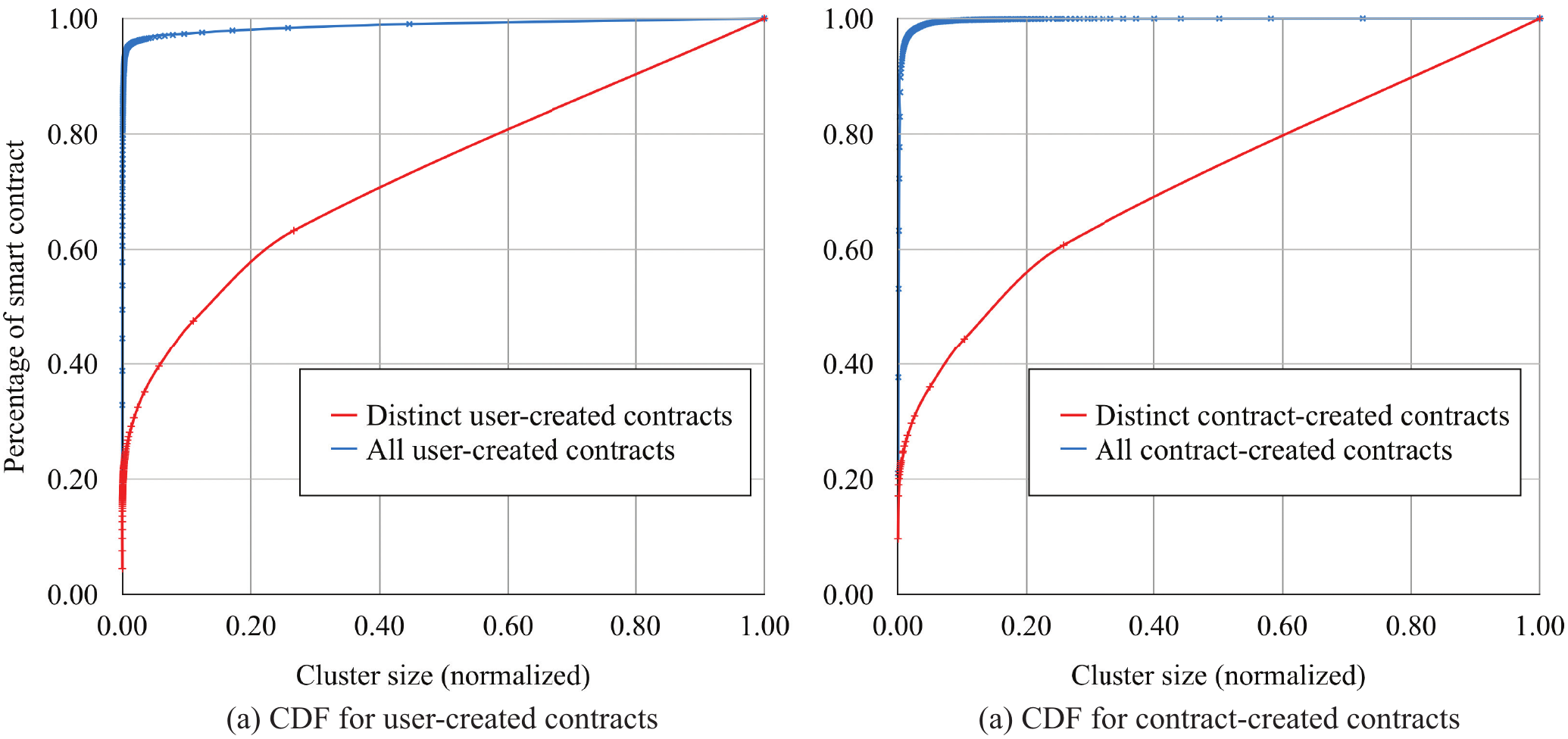}
\vspace{-0.15in}
\caption{CDF of smart contracts according to cluster sizes.}
\vspace{-0.2in}
\label{fig:cdf}
\end{figure}

We further investigate whether these clusters follow the \texttt{Pareto principle} (i.e., the 80/20 rule). The results suggest that the distribution of clusters follows a typical \texttt{Pareto Effect} after cluster size based normalization, as shown in Figure~\ref{fig:cdf}.
For the distinct contracts, the top 20\% of the clusters account for 60\% of the contracts. With regard to the whole contracts including all the duplicates, the top 1\% of the clusters account for over 95\% of the contracts (95.24\% and 95.30\% for the user-created and contract-created contracts, respectively).

\begin{table}[t]
\caption{Top 15 clusters for both user-created and contract-created contracts.}\vspace{-0.1in}
\label{cluster}
\resizebox{\textwidth}{!}{%
\begin{tabularx}{\textwidth}{cccc}
\hline
\multicolumn{2}{c}{User-created contracts} & \multicolumn{2}{c}{Contract-created contracts} \\
Size (with Duplicates) & Use & Size (with Duplicates) & Use \\ \hline\hline
3,338 (15,713) & ERC-20 token & 382 (1,799) & ERC-20 token \\
2,293 (19,263) & ERC-20 token & 295 (1,983) & ERC-20 token \\
1,596 (11,737) & ERC-20 token & 76 (1,210) & ERC-20 token \\
1,155 (6,174) & ERC-20 token & 43 (209) & ICO \\
1,022 (6,466) & ERC-20 token & 28 (223) & ICO \\
724 (4,494) & ERC-20 token & 20 (571) & Airdrop \\
662 (2,418) & ERC-20 token & 18 (39) & ERC-20 token \\
343 (1,054) & Other contract & 16 (30) & ITO \\
278 (972) & ERC-20 Token & 15 (922) & Airdrop \\
253 (509) & Fomo3D-like game & 13 (20) & Exchange wallet \\
228 (661) & ERC-20 Token & 12 (412) & User wallet \\
186 (342) & Other contract & 11 (67) & ICO \\
168 (343) & Other contract & 11 (229) & Airdrop \\
151 (281) & ERC-20 token & 11 (345) & Airdrop \\
143 (521) & ERC-20 token & 11 (2,581) & Airdrop \\
\hline
\end{tabularx}%
}
\vspace{-0.2in}
\end{table}

\noindent \textbf{What are these smart contract clusters?}
Table~\ref{cluster} lists the top 15 clusters for user-created contracts and contract-created contracts. We manually went through these clusters and labelled them according to their functionalities. Each type of contracts has its own characteristic functions, \textit{e.g.}, refund and deposit, airdrop and distribution, transfer and so on.
Our exploration suggests that the largest clusters mainly fall into the following categories:

\noindent \textbf{(1) ERC-20 Clusters.} 
ERC-20 related contracts take the majority of popular clusters.
We successfully identified a number of ERC-20 clusters, which might derive from different solc versions, as new versions of solc may bring in new opcodes; or more importantly, from different ERC-20 templates, as a result of different implementations of revisions of ERC-20 standard (\textit{e.g.}, the well-known OpenZeppelin~\cite{openzeppelin} libraries).

By manually analyzing the top 100 clusters, we have compiled a list of 53 different templates that were widely used in smart contracts.
Note that the similar contracts created by these templates were not necessarily plagiarized.

\noindent \textbf{(2) Game Contracts.}
Many popular clusters are game contracts.
The largest game cluster is \texttt{Fomo3D-like contracts}. 
Due to the popularity of \texttt{Fomo3D}, numerous developers just copied and pasted the original open-source contracts to create similar games.
Besides Fomo3D, other popular games such as \texttt{PoWH3D}~\cite{powh3d}
and \texttt{CryptoKitties}~\cite{CryptoKitties}, have contract clones as well.

\noindent \textbf{(3) ICO and AirDrop Contracts.}
ICO~\cite{ico} stands for Initial Coin Offering, the cryptocurrency equivalent of IPO (Initial Public Offering). It is a way for crypto startups to raise money by selling tokens. ICO has experienced an explosive growth since 2017 (and the bubble burst at the end of the third quarter 2018), which can be used to illustrate the phenomenon that a vast number of such contracts were deployed during this period. In terms of AirDrop contracts~\cite{airdrop}, attackers have to create massive such contracts to win the race of exploitation, which is a competition to defeat the flawed random function.

\noindent \textbf{(4) Other Contracts.}
We also observed that there do exist some short contracts with extremely simple operations, \textit{e.g.}, a pair of getter and setter, fetching data from storage, \textit{etc.}. Such contracts were grouped into clusters as well.

\noindent
\fbox{%
  \parbox{1\textwidth}{%
\vspace{-0.1in}
    \begin{center}
\emph{Observation-1: Although millions of contracts were deployed on Ethereum, most of them were duplicates and share same/similar code and functionalities, which suggested the homogeneous nature of the ecosystem.}
    \end{center}
\vspace{-0.1in}
  }%
}

\vspace{-0.1in}
\subsection{Vulnerability Provenance}
\vspace{-0.1in}

We then seek to explore the correlation between code clones and security vulnerabilities in two ways. First, we want to measure the vulnerability introduced by duplicate contracts, \textit{i.e.}, the original contracts are suffering from vulnerabilities, and other duplicate contracts (with same hash values) would inherit the vulnerabilities. 
Then, for the distinct contracts that were very similar, we seek to measure whether they have the same vulnerabilities introduced by code clones.

\noindent \textbf{Vulnerability Detection.}
To identify security vulnerabilities, we take advantage of a state-of-the-art tool~\cite{peckshieldscanner} developed by PackShield. It is a bytecode level static analysis framework composed of multiple program analysis techniques, including control flow analysis, data flow analysis and symbolic execution. We focus on 7 types of vulnerabilities that might cause damages with real impact, including (1) reentrancy, (2) overflow, (3) cross-function race condition, (4) mismatched constructor, (5) ownership takeover, (6) manipulable suicide address and (7) ERC-20 related vulnerabilities.
As it is not the emphasis of this paper, we will use the results directly without giving technical details of the tool.

\noindent \textbf{Vulnerable Duplicate Smart Contracts.}
We have scanned all the distinct contracts, inlcuding $74,647$ user-created and $3,964$ contract-created contracts.
It is interesting to see that, although only 25K distinct user-created contracts were vulnerable, considering all the duplicate contracts, we have identified over 1.2 million vulnerable contracts. As for the contract-created contracts, the result is more striking. Only 51 unique contract-created contracts were vulnerable, but we have identified over 2.2 million vulnerable contracts when we consider all the duplicates. This result suggests that a large number of duplicate contracts would suffer from the vulnerability issues inheriting from the original contracts. 

\noindent \textbf{Copy-paste Vulnerabilities.}
Then, we try to measure the copy-paste vulnerabilities from those similar contract pairs (with different hash values). For the 472K similar contract pairs we identified (with similarity scores over 70), we measure the similarity in vulnerabilities between them, \textit{i.e.}, whether they share the same types of vulnerabilities and the same number of vulnerabilities. 
\emph{For contracts that share both the same types and same number of vulnerabilities, we will mark them as having exactly the same vulnerability behaviors.}
Note that, we further differentiate the authors of the contracts to determine whether the contract pairs are code clones between different authors or the re-deployment from the same author. As shown in Table~\ref{vul_distribution}, we have classified the results into two general categories.

\textbf{Same Vulnerability Behaviors.}
Over 53\% of similar contract pairs have the same vulnerability behaviors. Over 46K contract pairs share the same vulnerabilities, and over 90\% of them were created by different authors. This indicates that when someone copied the code, he/she did not know that the original contracts were vulnerable, and thus inherited the same vulnerabilities.

\textbf{Different Vulnerability Behaviors.}
Over 46\% of the similar contract pairs have different vulnerability behaviors. It is interesting to see that, for over 149K contract pairs where only one contract is vulnerable, roughly 96\% of them were created by different authors. It indicates that when the authors copy and paste the code, they may have identified the vulnerabilities of the original contracts and thus patched them. Another scenario to explain this is that their modification of the original contracts may introduce new security vulnerabilities. Besides, over 12\% of the similar contract pairs were found sharing vulnerabilities, which could also be introduced by code reuse.

\noindent \textbf{Case Study.} Here, we use the Fomo3D-like game contracts as a case study. As we revealed in Section~\ref{subsec:cluster}, Fomo3D-like games were popular. We have identified 253 distinct contracts belonging to this cluster. As the original Fomo3d game suffers from the ``Airdrop Vulnerability'', over 80\% ($213$ out of $253$) of its contract clones also share the same vulnerability. 

\noindent
\fbox{%
  \parbox{1\textwidth}{%
\vspace{-0.1in}
    \begin{center}
\emph{Observation-2: Copy-paste vulnerabilities were prevalent in the smart contract ecosystem, duplicate contracts and similar contract would inherit security issues from the original vulnerable ones.}
    \end{center}
\vspace{-0.1in}
  }%
}

\begin{table}[t]
\caption{Distribution of vulnerability similarity across similar contract pairs.}
\vspace{-0.1in}
\label{vul_distribution}
\begin{tabularx}{\textwidth}{r|YY|YYY}
\hline
 & \multicolumn{2}{>{\centering\setlength\hsize{2\hsize} }X|}{Same vulnerability behaviors} & \multicolumn{3}{>{\centering\setlength\hsize{3\hsize} }X}{Different vulnerability behaviors} \\
 &\vspace{1ex}{Neither are vulnerable} & \vspace{1ex}{Both are vulnerable} & \vspace{1ex}{One is vulnearble} & Both are vulnerable and overlapped & Both are vulnerable but not overlapped \\ \hline\hline
Same author & 26,570 & 3,813 & 6,368 & 1,678 & 247 \\
Different author & 180,101 & 42,368 & 143,146 & 58,338 & 10,034 \\ \hline\hline
Total & 206,671 & 46,181 & 149,514 & 60,016 & 10,281 \\
\hline
\end{tabularx}
\vspace{-0.2in}
\end{table}
\vspace{-0.2in}

\subsection{Clone Detection of DApps}
\vspace{-0.1in}

DApps are increasingly popular in the Ethereum ecosystem.
As Ethereum DApps are usually open-source, the plagiaristic behaviors could also be widespread.
Different from the normal smart contracts, a DApp may consist of one or more smart contracts. To measure the extent of similarity between DApps, we proposed an advanced similarity detection method.

\noindent \textbf{Definition.}
Here, we use the term \emph{DApp Clones} to describe the scenario where two DApps deployed by different authors share the similar core functionalities.
We use the accounts to differentiate the authorship.
As a large number of smart contracts were created on top of templates, thus we will first eliminate the impact introduced by the templates based on the list we labelled in Section~\ref{subsec:cluster}.

\noindent \textbf{Approach.}
For a given DApp pair, we first construct a weighted bipartite graph for them, and conduct bipartite graph matching on the graph. 
A bipartite graph is a graph whose vertices (contracts) can be divided into two disjoint sets U and V, such that every edge connects a vertex in U to one in V, \textit{i.e.}, U and V are independent sets. 
Here, we will calculate the similarity score between contracts and take the score as the weight of the corresponding edge.
Specifically, we take advantage of the \textit{Kuhn–Munkres algorithm}~\cite{kmalgorithm} to identify the maximum matching - a set of the most edges with the following two properties: 1) no two edges share an endpoint; 2) the weight of edges must be guaranteed to be the highest. Therefore, we are able to calculate the similarity score between DApps with more than one contract. As the calculation is not commutative, \textit{i.e.}, $Sim(DApp1, DApp2) \neq Sim(DApp2, DApp1)$, we keep the higher one as the final score.

\begin{figure}[t]
\centering
\includegraphics[width=0.9\textwidth]{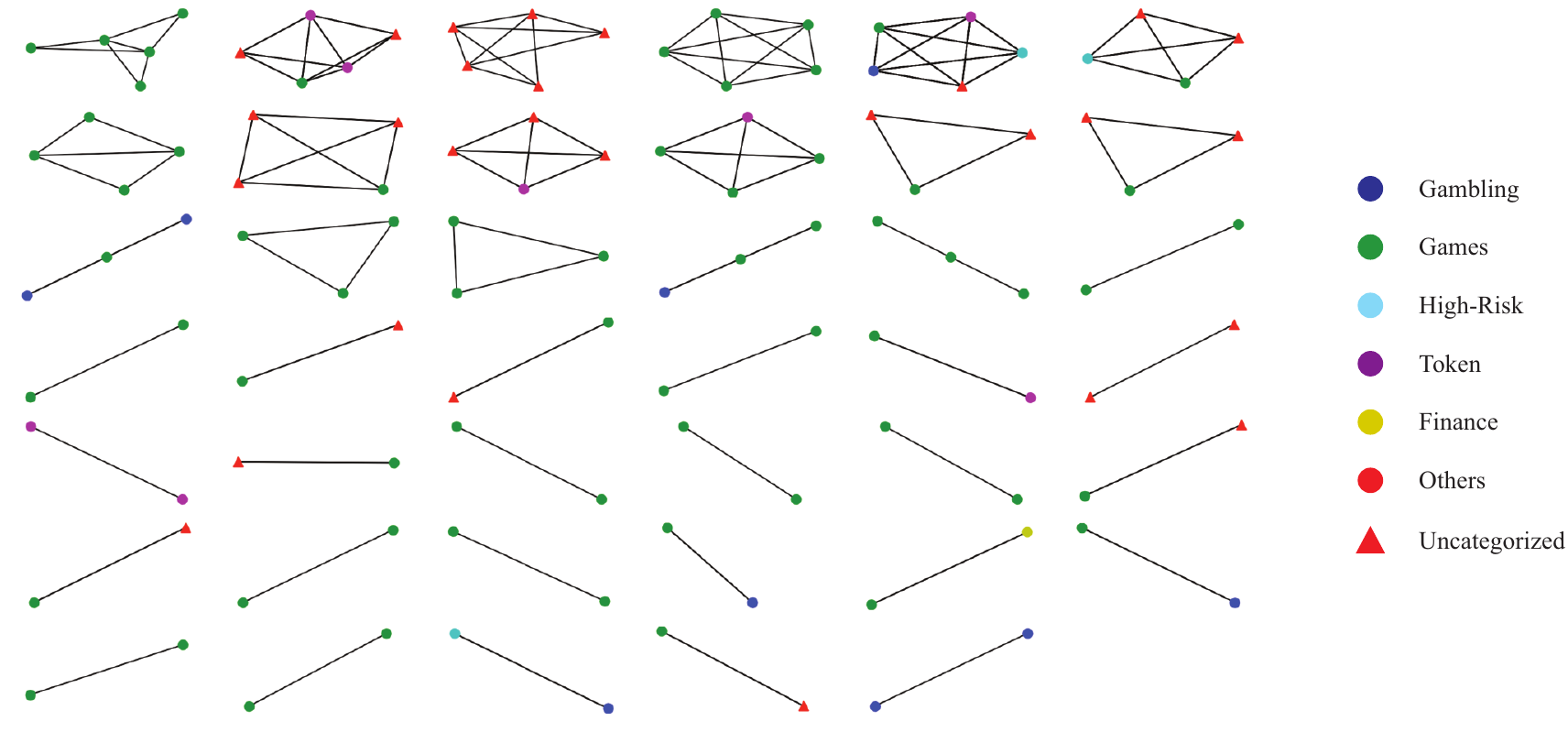}
\vspace{-0.2in}
\caption{Clustering results of 127 DApp clone pairs (114 unique DApps).} 
\vspace{-0.3in}
\label{fig:dapp_cluster}
\end{figure}

\noindent \textbf{Result.}
We have made our best efforts to collect $2,533$ DApps from several well-known DApp browsers, including DAppTotal~\cite{dapptotal}, DAppRadar~\cite{dappradar} and DAppReview~\cite{dappreview}. Note that we also crawled related metadata, \textit{e.g.}, category, volume, and the deployed time.
Based on the definition of DApp clones and using the above approach, we have successfully identified 127 DApp clone pairs with 114 distinct DApps in total. We further grouped them into 41 clusters by leveraging the approach mentioned in Section~\ref{subsec:cluster}. The results are shown in Figure~\ref{fig:dapp_cluster}. 

\noindent \textbf{Impact.}
To measure the impact of DApp Clones, we decided to take the historical volume as the indicator to identify the potential financial losses. Even worse, the high volume often means an active market which might attract more capital inflows~\cite{volume}, thus it would cause more damage to the original authors. 

In particular, we first analyzed all these 41 clusters and treated the earliest deployed DApp as the original one. 
Thus, we have 41 original DApps, and 73 plagiarized DApp clones in our dataset.
Then we calculated the differences between the original volume and the plagiarized volumes.

The overall volume of the 41 original DApps is $304,797.344$ ETH, while the volume of the 73 DApp clones reaches $89,565.321$ ETH, around 30\% of the original market.
The figures are diverse across the clusters by examining those clusters individually.
For 18 out of the 41 clusters, the volumes of the clones are higher than those of the corresponding original DApps, with some clones attracting two to three times more volumes than the original. In Table~\ref{table:plagiarism} (cf. Appendices), we summarized the statistics for the top 10 clusters in Figure~\ref{fig:dapp_cluster}.

\noindent
\fbox{%
  \parbox{1\textwidth}{%
\vspace{-0.1in}
    \begin{center}
\emph{Observation-3: DApp clones caused great financial losses to the original DApps, which exposed a contradiction between copyright protection and the open-source nature of the Ethereum ecosystem.}
    \end{center}
\vspace{-0.1in}
  }%
}

\vspace{-0.1in}
\section{RELATED WORK}
\vspace{-0.1in}

\noindent \textbf{Characterizing the Ethereum Ecosystem}
Several work have already been published to measure the Ethereum ecosystem~\cite{chen2018understanding, norvill2017automated,payette2017characterizing}. For example, Chen \textit{et al.} characterized money transfer, contract creation and contract invocation of Ethereum based on graph analysis~\cite{chen2018understanding}.
These studies may have a correlation with part of our work, however,
our work is the first systematic attempt to study contract clone phenomenon and its impact. Besides, some researchers focused on financial activities on Ethereum, including the Ponzi scheme~\cite{chen2018detecting} and ICO behavior~\cite{fenu2018ico}, which might be a complement to our work.

\noindent \textbf{Program Analysis of the Smart contracts}
Based on program analysis techniques (\textit{e.g.}, symbolic execution and formal verification), several frameworks have been proposed to detect security vulnerabilities in contracts~\cite{luu2016making, mythril, kalra2018zeus, tsankov2018securify}.
To the best of our knowledge, none of them performed a comprehensive study on the vulnerability provenance. 
In addition to vulnerability detection, 
some research studies were focused on topics including reverse engineering~\cite{zhou2018erays}, detecting gas-costly patterns~\cite{chen2017under}, automatically creating exploits~\cite{krupp2018teether}, etc.

\noindent \textbf{Code Clone Detection}
Code clone detection techniques have been studied
extensively for dozens of years, including text-based techniques~\cite{lee2005sdd, roy2008nicad}, token-based techniques~\cite{baker1995finding, baker1996parameterized, kamiya2002ccfinder, li2006cp}, counting-based techniques~\cite{yuan2011cmcd, yuan2012boreas}, and syntactic approaches~\cite{baxter1998clone, corazza2010tree, selim2010enhancing}, etc. 
These techniques were also widely explored in related domains, such as mobile app repackaging detection~\cite{crussell2012attack, crussell2013scalable, wang2015wukong, zhou2012detecting}.
In this work, we take advantage of a customized fuzzy hashing technique~\cite{fuzzyhashing}, which is both light-weight and effective.
Note that we did not rely on heavy-weight methods such as comparing the control-flow graph (CFG) and program dependency graph (PDG), mainly due to two reasons. 
First, our approach should be scalable. Second, the simplicity of smart contracts and EVM bytecode, \textit{i.e.}, the relatively simple logic and function invocations, makes it unnecessary to adopt those heavy approaches.
A limited number of studies have explored to study code clones in the smart contracts. For example, Kiffer \textit{et al}. identified substantial code reuse in Ethereum~\cite{kiffer2018analyzing}. Furthermore, Liu \textit{et al.} proposed ECLONE~\cite{liu2018eclone}, which is able to detect semantic clones for smart contracts. However, none of them have measured the ecosystem in large-scale, and characterized their security impacts. 

\vspace{-0.1in}
\section{Concluding Remarks and Future Work}
\vspace{-0.1in}

In this work, we present the first systematic attempt to characterize the code clone phenomenon in the Ethereum smart contract ecosystem. Based on the 10 million contracts dataset we harvested, we have revealed the homogeneity nature of the ecosystem. We also discovered and measured the security impacts of contract clones, for example, helping spread the security vulnerabilities and causing financial losses to the original DApps authors, etc.

There are a number of future lines of
work we will explore. First of all, we empirically use $70$ as the threshold to calculate the similarity scores, which can be improved by adopting adaptive approaches. Secondly, we may have coverage issues on manually labelling the contract templates, which can be alleviated by exploring some machine learning techniques. Lastly, part of our findings, such as those economic intensive phenomena in the Ethereum ecosystem, deserve more focused studies. Nonetheless, we believe our efforts and observations could positively contribute to the community and promote the best operational practices for smart contract developers.


\newpage
\section*{Appendices}
\label{sec:appendix}

\subsection{Fingerprint Generation Algorithm}

The detailed fingerprint generation algorithm is shown in Algorithm~\ref{algo:fingerprintalgo}.

\begin{algorithm}[h]
\caption{Generating the fingerprint for smart contract.}
\label{algo:fingerprintalgo}
\textbf{Input:} \textit{bytecode} of arbitrary contract\\
\textbf{Output:} Fingerprint \textit{fp}\\
\textbf{Description:} \textit{pc} - character representing current piece, \textit{ph} - the piece hash, \textit{tv} - trigger value, \textit{b64map} - mapping integer to base64 character
\begin{algorithmic}[1]

\Procedure{GenerateFp}{$bytecode$}
    \State InitTriggerValue($tv$)
    \State InitBase64Map($b64map$)
    \State InitPieceCharacter($pc$)
    \State InitPieceHash($ph$)
    \State $pieces \leftarrow$ CutOff($bytecode, tv$)
    \ForAll{$piece$ from $pieces$}
        \State UpdatePieceHash($ph, piece$)
        \State MapToPieceCharacter($pc, ph, b64map$)
        \State $fp \leftarrow$ Concatenate($fp, pc$)
        \State InitPieceHash($ph$)
    \EndFor
    \State \textbf{return} $fp$
\EndProcedure

\end{algorithmic}
\end{algorithm}

\subsection{Top Dapp Clone Clusters and Their Volumes}

Table~\ref{table:plagiarism} lists the statistics of the top 10 Dapp clone clusters.

\begin{table}[h]
\caption{Top 10 Dapp Clone clusters and their volumes (ETH).}
\label{table:plagiarism}
\begin{tabularx}{\textwidth}{ccccc}
\hline
Original DApp & \# Clones & Original volume & Plagiarized volume & Ratio \\
\hline\hline
CryptoCountries~\cite{cryptocountry} & 4 & 67,885.244 & 2.355 & \textless{}0.01\% \\
PoWTF~\cite{powtf} & 4 & 331.074 & 1,012.649 & 305.87\% \\
Po50~\cite{po50} & 4 & 76.801 & 213.058 & 277.42\% \\
Pepe Farm~\cite{pepefarm} & 4 & 25.428 & 33.577 & 132.05\% \\
Crypto Miner~\cite{minertoken} & 4 & 17,312.026 & 155.437 & 0.90\% \\
PoWH 3D~\cite{powh3d} & 4 & 187950.872 & 1778.146 & 9.38\% \\
CryptoTubers~\cite{cryptotubers} & 3 & 95.378 & 470.967 & 493.79\% \\
PoHD~\cite{pohd} & 3 & 242.607 & 5867.961 & 2418.71\% \\
Proof Of Craig Grant Coin~\cite{pocg} & 3 & 642.056 & 94.315 & 14.69\% \\
Crypto Gaming Coin~\cite{cryptogamingcoin} & 3 & 4.711 & 555.142 & 11783.95\% \\
\hline
\end{tabularx}
\end{table}

\end{document}